\documentclass[sigplan,nonacm]{acmart}
\usepackage{amsmath,amsfonts}
\usepackage{algorithmic}
\usepackage{graphicx}
\usepackage{textcomp}
\usepackage{xcolor}
\usepackage{braket}
\usepackage{multirow}
\usepackage{enumitem}
\usepackage{svg}

\usepackage{graphicx}
\usepackage{subcaption}

\newcommand{\ours}{{\sc{GANDALF }}}
\newcommand{\oursnspace}{{\sc{GANDALF}}}

\begin{document}

\title{Enabling Fast and Accurate Neutral Atom Readout through Image Denoising} 


\author{Chaithanya Naik Mude}
\affiliation{%
  \institution{ Department of Computer Sciences, University of Wisconsin–Madison}
  \city{Madison}
  \state{WI}
  \country{USA}
}
\author{Linipun Phuttitarn}
\affiliation{%
  \institution{Department of Physics, University of Wisconsin–Madison}
  \city{Madison}
  \state{WI}
  \country{USA}
}
\author{Satvik Maurya}
\affiliation{%
  \institution{Department of Computer Sciences, University of Wisconsin–Madison}
  \city{Madison}
  \state{WI}
  \country{USA}
}
\author{Kunal Sinha}
\affiliation{%
  \institution{Department of Physics, University of Wisconsin–Madison}
  \city{Madison}
  \state{WI}
  \country{USA}
}
\author{Mark Saffman}
\affiliation{%
  \institution{Department of Physics, University of Wisconsin–Madison}
  \city{Madison}
  \state{WI}
  \country{USA}
}
\affiliation{%
  \institution{Infleqtion}
  \city{Madison}
  \state{WI}
  \country{USA}
}

\author{Swamit Tannu}
\affiliation{%
  \institution{ Department of Computer Sciences, University of Wisconsin–Madison}
  \city{Madison}
  \state{WI}
  \country{USA}
}






\begin{abstract}

Neutral atom quantum computers hold promise for scaling up to hundreds of thousands or more qubits, but their progress is constrained by slow qubit readout. Parallel measurement of qubit arrays currently takes milliseconds,—much longer than the underlying quantum gate operations—making readout the primary bottleneck in deploying quantum error correction. Because each round of QEC depends on measurement, long readout times increase cycle duration and slow down program execution.

Reducing the readout duration speeds up cycles and reduces decoherence errors that accumulate while qubits idle, but it also lowers the number of collected photons, making measurements noisier and more error-prone. This tradeoff leaves neutral atom systems stuck between slow but accurate readout and fast but unreliable readout.

We show that image denoising can resolve this tension. Our framework, \oursnspace, uses explicit denoising using image translation to reconstruct clear signals from short, low-photon measurements, enabling reliable classification at up to $1.6\times$ shorter readout times. Combined with lightweight classifiers and a pipelined readout design, our approach both reduces logical error rate by up to $35\times$ and overall QEC cycle time up to $1.77\times$ compared to state-of-the-art convolutional neural network (CNN)-based  readout for Cesium (Cs) Neutral Atom arrays.
\end{abstract}

\maketitle 

\section{Introduction}







Neutral atoms quantum computers have emerged as a promising platform for building scalable quantum computers. In recent years, they have demonstrated the ability to trap and control thousands of individual atoms---with systems reaching up to 6,100 qubits ~\cite{Manetsch_2025, Chiu_2025}. These systems are also making progress towards fault tolerance~\cite{sales2025experimental, Bedalov2024}, with successful demonstrations of Quantum Error Correction~(QEC)\cite{bluvstein2024logical}, and are experimentally demonstrating applications in areas such as analog quantum simulation\cite{bernien2017probing}, and optimization~\cite{graham2022multi}.

Despite these advancements, a major challenge for neutral atom platforms is their relatively slow operation speeds. Table~\ref{tab:ops} summarizes the bottlenecks in QEC cycle---including two-qubit gates, readout, and fraction of readout in full quantum error correction cycles for leading quantum error correction codes---for the neutral atom systems~\cite{Wintersperger_2023}. For reference, we also show the operational latency for Google's superconducting hardware~\cite{google2025}.

A large fraction of the total time during fault-tolerant quantum computing is spent on readout, as \textit{qubit readout is the slowest and most error-prone step}, which not only limits the speed but also significantly impacts the error suppression capabilities of the quantum error correction code.



\textbf{Why is readout slow in neutral atom architectures?} In neutral atom systems, all atoms in the array are illuminated with a laser. If a qubit is in state $\ket{1}$, it will emit photons; if it is in state $\ket{0}$, it will remain dark. These scattered photons are collected by a highly sensitive camera to determine each qubit's state. However, each atom can at most emit a fixed number of photons per unit time, and the optics can only collect and detect a small fraction of the emitted photons.

\begin{table}[t]
\label{tab:latency_comp}
\centering
\small
\caption{Operational latencies for superconducting~\cite{google2025} versus neutral-atom quantum processors~\cite{Wintersperger_2023}.}
\label{tab:ops}
\begin{tabular}{lcccc}
\hline
\textbf{Hardware} & \textbf{Two Qubit} & \textbf{Readout} & \textbf{Readout Fraction}  \\
\textbf{Platform} & \textbf{Latency} & \textbf{Latency} & \textbf{in QEC Cycle}  \\
\hline 
Google~\cite{google2025} & 34~ns & 500~ns & $\sim$0.5 \\
Neutral Atom~\cite{Wintersperger_2023}  & 400~ns & 1~ms & $\sim$0.75 \\
\hline 
\end{tabular} 
\end{table}

\begin{figure*}[ht]
    \centering
    \includegraphics[width=0.92\linewidth]{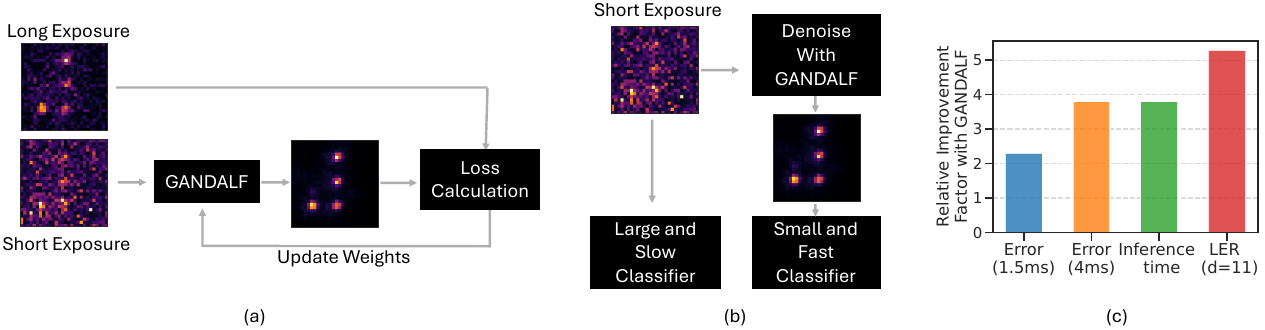}
    \caption{(a) Overview of training GANDALF to generate denoised images close to high-signal-to-noise ratio (SNR) long exposure images from low-SNR short exposure images. (b) The baseline classifier~\cite{phuttitarn2023enhanced} is trained on noisy short-exposure images using ground-truth labels extracted from long-exposure images, whereas denoising with GANDALF enables a small, fast classifier to achieve similar accuracy. (c) Relative improvement factor compared to baseline design that uses deep CNN~\cite{phuttitarn2023enhanced} in readout fidelity (at readout durations $1.5~\rm ms$, $4~\rm ms$), inference time, and logical error rate (LER) for surface code ($d=11$).}.
    \label{fig:GAN}
    \vspace{-0.1in}
\end{figure*}


To measure a qubit accurately, the system needs to collect a sizable number of photons from the atom. To increase the collected photons, we can increase the laser power or increase the duration of the readout pulse. 
Increasing the power only increases the photon emission rate up to a limit set by atomic parameters, and risks overheating the atom, which can cause it to escape its trap, or jump to an unwanted energy level, or experience crosstalk. The other option is to stretch the readout pulse, increasing exposure time to collect more photons, which slows down readout. As a result, designers face a trade-off: faster readout collect fewer photons and are less reliable, while slower readout improve accuracy but take more time.

Ideally, we want fast readout without using high laser power, which translates to performing classification at significantly lower photon counts. 
Most existing readout classifiers~\cite{phuttitarn2023enhanced, kent2025efficientmeasurementneutralatomqubits} rely on high photon counts to distinguish between qubit states. Higher photon counts create a clear separation between $\ket{0}$ and $\ket{1}$ states. But this assumption breaks down for fast, low-photon-count measurements, because it's hard to distinguish state-dependent illumination of an atomic qubit from background noise caused by spurious photons.

We observe that the existing classifiers struggle to maintain the same level of accuracy at low photon count due to low signal to noise (SNR) ratio. This limitation presents a significant bottleneck for enhancing quantum error correction, calibration, and application-level performance.

In this work, we demonstrate that denoising can enable high-fidelity readout  at low photon counts and can be effectively used for accelerating the readout for neutral atom quantum computers. We utilize a dataset of paired fluorescence images from a $3\times3$ Cesium (Cs) atom array, where a high \textit{Signal to Noise Ratio (SNR)} path provides long-exposure ground truth labels and an attenuated path produces noisy, short-exposure measurements~\cite{phuttitarn2023enhanced}. We observe that conventional CNN classifiers, at low photon counts, can not differentiate the background noise from qubit-emitting photons. We believe neutral-atom images lack semantic richness of natural images, where typical CNN based computer vision model have shown significant accuracy even under noisy conditions. This motivates a different approach of processing the noisy images prior to classification to improve the SNR by filtering the noise, instead of classifying them directly.



Our central design insight is to decouple signal recovery from classification.
We introduce a Generative Adversarial Network for Detection at Low Fluorescence levels (GANDALF), a generative adversarial framework that \textit{denoises low-photon images, restoring per-site fidelity and suppressing photon noise and crosstalk improving SNR.} By training on paired datasets of low- and high-photon fluorescence frames from cesium atom arrays~\cite{phuttitarn2023enhanced}, our approach reconstructs high-fidelity images from fast, low-photon measurements—enabling rapid and accurate qubit state classification. Figure~\ref{fig:GAN} shows that GAN can significantly improve the signal-to-noise ratio for a short-exposure atom array image, enabling high-fidelity readout even at low photon counts. 

Moreover, with denoising, amplifying the signal, the classification task becomes far easier, enabling the use of lightweight site-specific classifier models such as feedforward networks, matched-filter networks, or their hybrids \cite{kent2025efficientmeasurementneutralatomqubits}. These compact classifiers match the accuracy of larger  CNNs~\cite{phuttitarn2023enhanced} while cutting latency by up to five times and parameter count by more than an order of magnitude, making them practical for arrays with thousands of qubits. Furthermore, we propose a readout architecture with a pipelined denoising and classification, so that they are not in the critical path. This opens the door to using sophisticated ML denoising and classification algorithms running on classical accelerators~\cite{nvidia_nvqlink}.     

Finally, by making the denoising model fully convolutional, the same network trained on small calibration arrays seamlessly generalizes to larger lattices, maintaining a nearly constant per-site inference cost as the system size grows. This is crucial as generating training data for fully occupied qubit lattices can be slow experimentally due to slow loading of atoms on current systems. 
Our evaluations show that this combination of denoising with GAN and a lightweight classifier, such as a shallow Feed Forward Neural Network (FNN), delivers significant gains. At exposure times of $1.5~\rm  ms$, GANDALF reduces readout error by $2.8\times$ compared to CNN baselines~\cite{phuttitarn2023enhanced}. The dataset we use is imperfect, and we expect higher-quality data to be available in the near future. Therefore, to understand the best-case improvements bu using denoising, we create a dataset generated via confidence-based filtering and observe sub-1\% inaccuracy even at the shortest duration of $1.5~\rm ms$. This highlights the effectiveness of our approach in reducing readout duration close to millisecond region for Cesium~(Cs) atom arrays.  

At the system level, logical error rates improve dramatically—up to $35\times$ reduction for bivariate bicycle codes and $5\times$ for surface codes—demonstrating that faster, denoised readout directly strengthens fault tolerance. Lightweight classifiers achieve inference in the 100 $\mu$s range, and pipelining shortens overall QEC cycle time by $1.77\times$, reducing idling-induced decoherence. Even with commodity GPUs with only 4096 CUDA cores, the denoiser scales from $3\times3$ to $64\times64$ and beyond, with sublinear latency growth, ensuring scalability for next-generation atom-array processors.

Our work is the first to overcome the low-SNR barrier in neutral-atom readout by reconstructing high-SNR readout images prior to classification. This architectural change shifts the burden of readout accuracy from hardware factors, such as exposure time and laser power, to an upstream denoising stage, enabling accurate, fast readout and substantially lower logical error rates, as well as shorter QEC cycle times. Our design scales naturally to large atom arrays and integrates seamlessly into efforts on connecting neutral atom quantum platforms with GPUs for scaling QEC~\cite{nvidia_nvqlink,caldwell2025platform}. This paper makes the following contributions:


\begin{itemize}[topsep=0pt, leftmargin=*]

\item We demonstrate that denoising enables reliable state discrimination at much shorter exposure times, making low-photon and fast readout viable for fault-tolerant execution.

\item We show that generative denoising adapted to the non-semantic, site-local structure of neutral-atom readout images yields higher classification accuracy than existing downstream methods operating directly on raw readout images.

\item We demonstrate scalability with our fully convolutional denoiser whose learned filters apply consistently from small arrays to large atom lattices without architectural changes, maintaining per-atom-site readout accuracy.

\item We show \ours reduces readout error by up to $2.5\times$, lowers logical error rates up to $5\times$ for surface code and $35\times$ for the bivariate-bicycle (BB) code, and shortens quantum error correction cycle time by $1.77\times$, directly addressing readout bottlenecks in neutral atom quantum computers.

\end{itemize}

\section{Background}

\subsection{Qubits and gates for neutral atom quantum computing}

Neutral atom platforms have emerged as a promising architecture for scalable quantum computing, offering flexible qubit layouts, long coherence times, and high connectivity~\cite{Wintersperger_2023,saffman2019quantum}. In these systems, individual atoms—typically species such as Rubidium~(Rb), Cesium~(Cs), Strontium~(Sr), or Ytterbium~(Yb)—serve as qubits, with internal states encoding the computational $\ket{0}$ and $\ket{1}$ states.

Qubits are held in space using optical tweezers\footnote{Arrays of optical tweezers are generated using acousto-optic deflectors~(AOD) or spatial light modulators~(SLM), allowing dynamic reconfiguration and parallel operation.}, the tightly focused laser beams, that create potential wells to trap single atoms. The vacuum environment and tight localization of atoms enable long coherence times, while the high-numerical-aperture lenses allow for the precise optical control and measurement. Single-qubit gates are implemented\footnote{In recent experiments~\cite{madjarov2020highfidelity,Wintersperger_2023}, these operations are performed in parallel across the array, with gate fidelities exceeding 99.5\%.} using resonant laser or microwave pulses that drive coherent rotations between the two qubit states. Two-qubit entangling gates are typically realized via the Rydberg blockade mechanism\footnote{Atoms are excited to high-energy Rydberg states where strong dipole-dipole interactions prevent simultaneous excitation of nearby qubits.} which prevents simultaneous excitation of nearby qubits~\cite{isenhower2010demonstration, levine2019parallel}. 

These atomic interactions enable fast, high-fidelity two-qubit gates with interaction ranges spanning several microns, making neutral atom systems especially well-suited for two-dimensional lattice-based quantum error correction codes and programmable quantum simulations.


\subsection{Readout in neutral atom architecture}

In neutral atom quantum computers, qubit states are typically encoded in long-lived hyperfine or electronic levels and readout relies on detecting whether the atom is in a particular internal state. Readout works by repeatedly shining a laser tuned to excite only one of the qubit states. If the atom is in the excited state (e.g., $\ket{1}$), it will scatter photons and appear bright under a camera. If it is in the other computational state (e.g., $\ket{0}$), it remains dark. By capturing this fluorescence with a sensitive camera, the system determines the atom’s state.



Although conceptually simple, this readout process is slow compared to the speed of quantum gates. To confidently distinguish between bright and dark states, the system must collect enough photons to overcome background and camera noise. This requires either using a stronger laser or increasing the exposure time. A stronger laser can cause heating, which may knock the atom out of its trap. Conversely, a longer exposure slows the readout and limits the rate at which quantum circuits can be executed.

Additionally, the fidelity of readout depends on many experimental parameters, including the numerical aperture of the imaging system, the qubit species, and spacing between atoms.

\begin{figure}[!t]
\centering
\includegraphics[width=0.8\columnwidth]{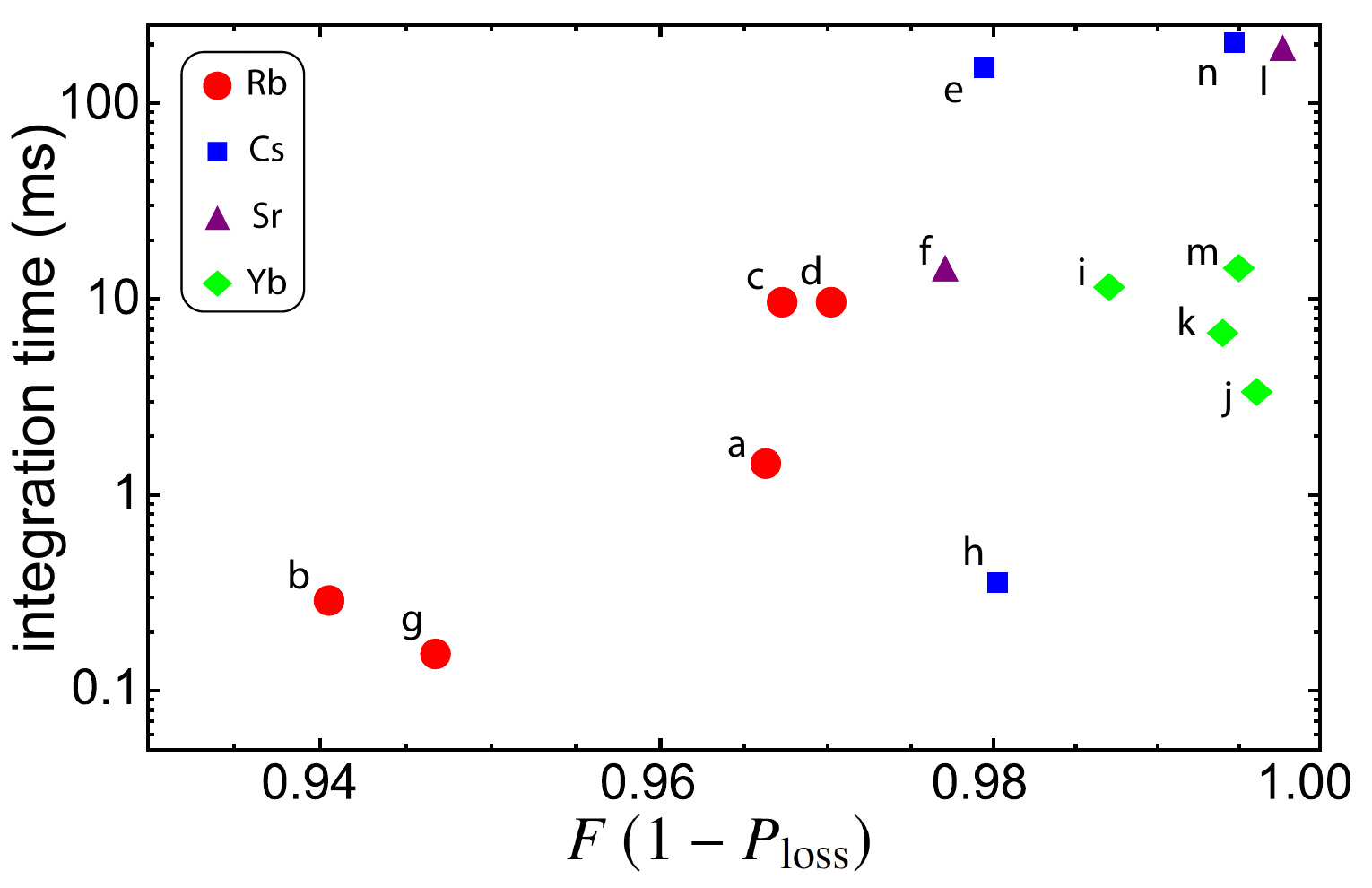}
\caption{Representative measurements of neutral atom qubit states in free space reported in the literature. The results are quantified in terms of the measurement time and the generalized fidelity given by the state detection fidelity $F$ times the atom retention probability $1-P_{\rm loss}$ . The experiments a,b,g,h  with the shortest integration times used single photon counting detectors and all others used cameras:  
a \cite{Fig2a},
b \cite{Fig2b},
c \cite{Fig2c},
d \cite{Fig2d},
e \cite{Fig2e},
f \cite{Fig2f},
g \cite{Fig2g},
h \cite{Fig2h},
i \cite{Fig2i},
j \cite{Fig2j},
k \cite{Fig2k},
l \cite{Fig2L},
m \cite{Fig2M},
n \cite{scott2025lasercoolingqubitmeasurements}. We refer the readers to \cite{saffman2025quantum} for additional discussion.}
\label{fig:time_distribution}
\end{figure}

\textbf{Effect of choice of atomic species on fast readout.}
Qubit readout remains a key bottleneck in neutral atom quantum computing, with significant variations in fidelity and speed across available atomic species, including Cs, Rb, Yb, and Sr illustrated in Figure~\ref{fig:time_distribution}. The performance of fluorescence-based state detection depends strongly on species-specific properties such as cycling transition strength, background scattering, and susceptibility to state leakage. Each atomic species imposes distinct system-level constraints such as photon budget, trap spacing, and detection optics, necessitating a careful balance between fidelity, speed, complexity, and scalability. \footnote{We note that a readout time of $1.5~\rm ms$ is used in this work as an example of short duration measurements. In practice the lower limit on atom measurement time depends on many experimental parameters including the specific atom and atomic transitions used, optical detection system efficiency, camera sensitivity, and noise. Advanced techniques suggest that it should be possible to acquire high SNR images in $\sim 100~\mu\rm s$ \cite{scott2025lasercoolingqubitmeasurements}. Adding high performance image processing as described here has the potential to further reduce the effective measurement time.


}

\subsection{Why does fast readout matter?}
Readout is essential for extracting useful information from a quantum computer, its speed and reliability pose a major challenge for scaling up neutral atom systems and supporting fast feedback in quantum error correction. The execution time of a fault-tolerant quantum program can be coarsely approximated based on logical operations, QEC cycle repetitions and time taken by each QEC cycle, as,
\begin{equation}
\small
\text{Execution time} \approx
(\#\text{T-gates}) \times (d\text{-rounds}) \times (\text{QEC cycle time})
\end{equation}

Each component in this model encapsulates a key bottleneck to the scalable, fault-tolerant quantum computation,

\begin{itemize}[topsep=0pt, leftmargin=*]
\item Non-Clifford Gates (e.g., $T$ gates) are necessary for universal quantum computation. They are the slowest operations due to magic state distillation~\cite{gidney2025factor}. 
\item $d$-Rounds represent the number of QEC iterations needed to suppress logical errors for a code of distance $d$. As expected, larger $d$ values are required for higher physical error rates, increasing overall runtime.
\item QEC cycle time is the duration of one round of quantum error correction, where parity checks (syndromes) are generated to track errors and keep logical qubits free of errors.
\end{itemize}

We observe, for most qubit technologies the duration of each QEC cycle is fundamentally bounded by the time it takes to measure qubits and reinitialize them—two steps that are dominated by readout latency in neutral atom systems:
\begin{equation}
\small
\text{QEC Cycle Time} = \text{Syndrome Generation} + \text{Readout} + \text{Reset}
\end{equation}

While this model emphasizes the linear relationship between readout latency and QEC cycle duration, it omits other crucial—but often overlooked—roles that fast readout plays in the performance of neutral atom quantum computers:

\textbf{Atom loading and rearrangement overhead.} In neutral atom platforms, building a defect-free 2D qubit lattice requires stochastic loading\footnote{This initialization process is slow, imaging intensive, and requires repeated fluorescence-based measurements to identify occupied sites and reposition atoms using optical tweezers.} followed by iterative rearrangement, or near-deterministic loading from a reservoir which nevertheless requires a verification imaging step to achieve perfect array filling. \cite{Norcia2024,Gyger2024}. Fast, high-fidelity readout accelerates each stage of this pipeline—enabling denser lattices, reducing the probability of atom loss during relocation, and improving experiment yield. Though often hidden behind “setup time,” inefficiencies here can dominate runtime in large-scale systems or in hybrid pipelines where reloading is triggered adaptively.

\textbf{QEC bootstrapping and syndrome verification.} For QEC codes to start correctly, not only initialization, but ancilla qubits must be verified through measurement. Faster readout enables quicker feedback, reducing dead time and improving the throughput of syndrome generation across the code lattice.


\textbf{Pipelining and reuse of physical infrastructure.} In principle, multiple Magneto-Optical Traps (MOTs) with reloading tweezers \cite{Norcia2024, Gyger2024} can be deployed in parallel to mask atom loading time, but doing so increases system complexity, cost, and physical footprint. Faster readout alleviates the need for deep pipelining by making each atom loading pass more productive, reducing the amortized cost per qubit and easing scalability constraints.

\section{Understanding challenges in fast readout}
Readout speed directly governs QEC cycle time, its influence is far broader—impacting lattice preparation, code startup, and system utilization efficiency. Fast readout is not just a lever for shorter execution times; it is a key enabler for practical, reliable, and reconfigurable fault-tolerant quantum computing on neutral atom platforms. In this work, we explore the advantages of employing machine learning to improve the accuracy and speed of measuring atomic qubits.


\subsection{Experimental setup and dataset}
We utilize the dataset generated by the neutral atom system consisting of $3\times3$ grid of cesium atoms, each held in place by optical tweezers as mentioned in prior work~\cite{phuttitarn2023enhanced}. To read out the state of each atom, we use a laser  beam that causes atoms in the “bright” state to fluorescence, while the “dark” state atoms emit few to no photons. The emitted photons are then captured by a sensitive camera through two optical paths: (1) a bright, high-quality path used for generating accurate labels, and (2) a dimmer, noisier path used for generating readout images for testing machine learning models. 

\begin{figure}[!t]
    \centering
    \includegraphics[width=\linewidth]{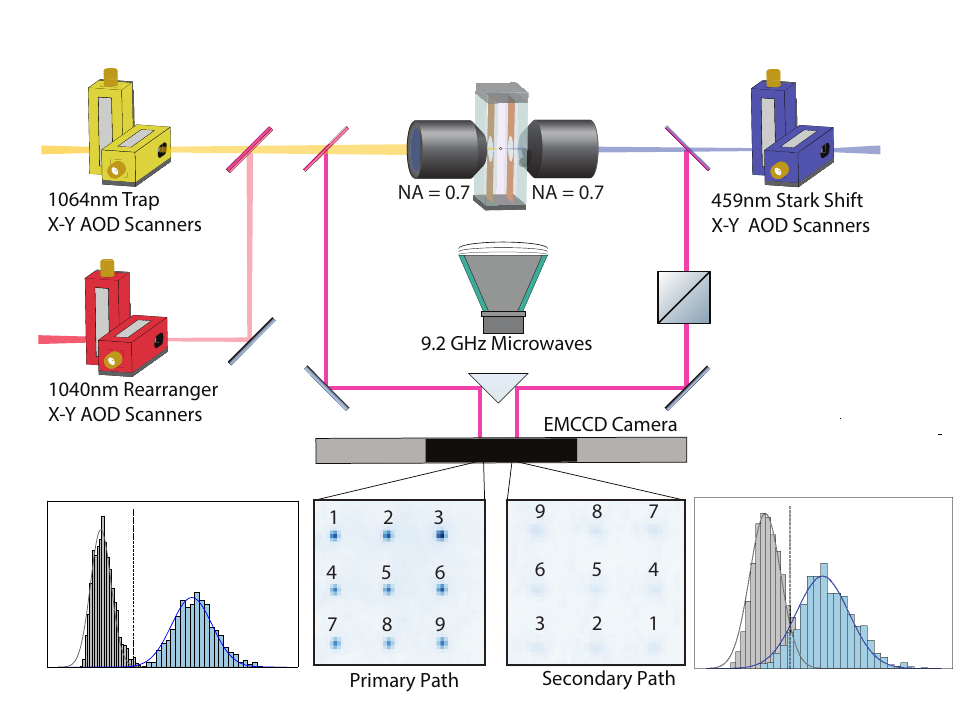}
    \caption{The experimental setup~\cite{phuttitarn2023enhanced} uses optical tweezers to trap the atoms and an EMCCD camera to capture the photons emitted by the atoms due to fluorescence. The primary path captures the entire signal, i.e, photons emitted by the atom during the fluorescence, while the secondary path attenuates this signal by $10\times$ resulting in an equivalent shorter exposure setting, corresponding to $10 \times$ reduction in readout time. The difference in captured photons of primary and secondary optical paths is evident from the histograms with better separation for histograms of the long exposure~(primary) images relative to the short exposure~(secondary) images.}
    \label{fig:experimental setup}
\end{figure}

The experimental setup is shown in Figure~\ref{fig:experimental setup} is used to collected the $3000$ readout images of $3\times3$ grid of Cesium (Cs) atom arrays resulting in $27000$ single-site images, with exposure durations ranging from $15~\rm ms$ to $100~\rm ms$ for primary path and from $1.5~\rm ms$ to $10~\rm ms$ for secondary path with noisy images. The atom presence labels are generated using primary path using thresholding methods and the noisy images are used for testing machine learning models.


\begin{figure}[t]
\centering
\includegraphics[width=\columnwidth]{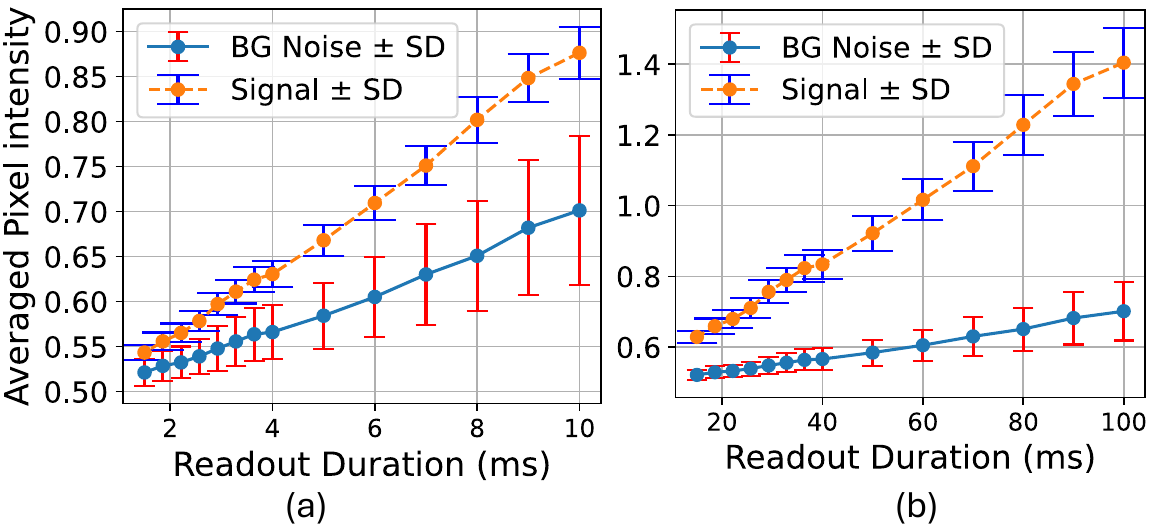}
\caption{Comparison of pixel intensity when atom is present, i.e, signal strength, and when atom is absent, i.e, background noise (BG Noise) for (a) shorter exposure time $1.5~\rm ms$ to $10~\rm ms$ with secondary path images, and (b) longer exposure time $15~\rm ms$ to $100~\rm ms$ with primary path images.}
\label{fig:compare_short_long_readout}
\end{figure}

\subsection{Low photon counts limits fast readout}

Each readout image obtained by the EMCCD camera consists of fluorescence from atoms arranged on a fixed lattice, where each site is either bright ($\ket{1}$) or dark ($\ket{0}$). Fast neutral-atom readout is fundamentally limited by the number of fluorescence photons collected during a measurement. Photon arrivals follow Poisson statistics,\footnote{The number of detected photons in a given exposure has mean $N$ and variance $N$, yielding a fundamental shot-noise level of $\sqrt{N}$.} so, while the signal grows linearly with exposure time~$\tau$, the signal-to-noise ratio (SNR) improves only as $\sqrt{\tau}$. As shown in Figure~\ref{fig:compare_short_long_readout}, short exposures (1.5--10\,ms) collect far fewer photons and produce signals that barely exceed the background noise floor, whereas longer exposures (15--100\,ms) accumulate enough photons to clearly separate occupied and unoccupied sites. This scaling of signal-to-noise ratio makes aggressive reductions in exposure time rendering fast readout inherently challenging.

\begin{figure}[t]
\centering
\includegraphics[width=\columnwidth]{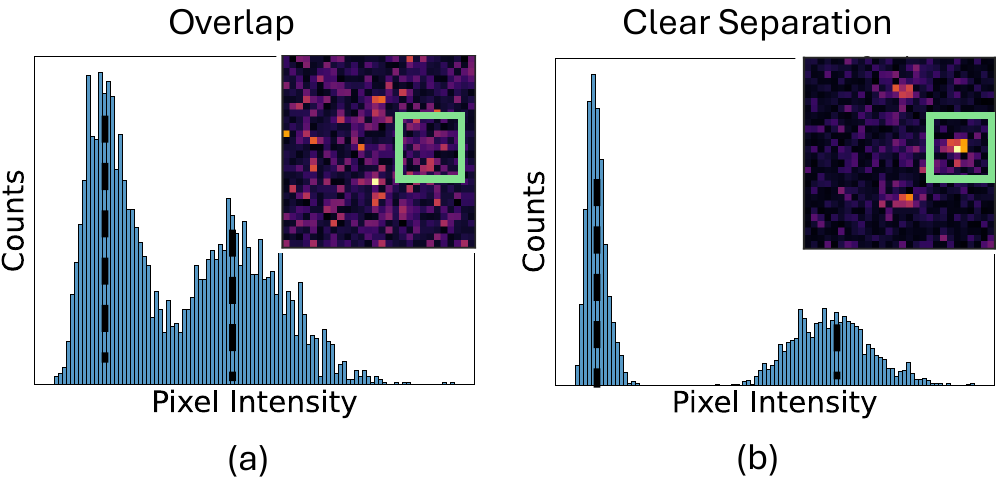}
\caption{ Visualization of histogram of pixel intensity of the atom site over all the images for (a) short exposure attenuated secondary path images and (b) long exposure images primary path images. The presence of significant overlap of histograms for short exposure between the two histograms pose a significant challenge for conventional threshold-based classification.}
\label{fig:site-averaged_noisy}
\end{figure}


\subsection{Downstream classification challenges}
The reduction in photon counts at short exposures also significantly complicates downstream state classification. As shown in Figure~\ref{fig:site-averaged_noisy}(b), long-exposure images yield well-separated intensity distributions for bright and dark sites, allowing simple classifiers, including thresholding with Gaussian or square masks and Gaussian mixture models~(GMMs), to perform reliably. At short exposures, however, the distributions for the two states overlap substantially due to insufficient photon counts as shown in Figure~\ref{fig:site-averaged_noisy}(a), eliminating the contrast required for these lightweight methods to discriminate states accurately. This loss of separability necessitates more expressive downstream machine learning classifiers capable of extracting discriminative features from weak, noise-dominated short exposure readout images.


Prior work~\cite{phuttitarn2023enhanced} developed two CNN-based classifiers that substantially improve readout accuracy relative to thresholding approaches. \textbf{These CNNs are trained and tested on the low-exposure noisy images with the labels generated from the high-quality optical path.} CNN-site (355k parameters) operates on individual lattice sites and reduces classification errors by up to 56\% in low-crosstalk settings, while CNN-array (74M parameters) incorporates neighboring-site information and improves error rates by up to 43\% and crosstalk misclassification by 78\% compared to Gaussian thresholding.

Although these models meaningfully extend the feasible operating range to moderately shorter exposures, their accuracy remains limited when photon counts are very low, as they must classify directly from heavily degraded images with minimal signal contrast. Further gains would require even larger and more complex CNNs, an approach that is both ineffective—because the underlying signal is too weak—and impractical due to the growing model size, resulting in long inference times for larger arrays. This necessitates   a more scalable strategy that alleviates the burden on classifiers.

\subsection{Why are standard CNN models not enough?} CNNs are the de-facto models for image classification. In natural-image domains with rich semantic content and large training datasets, they perform remarkably well, even in the presence of moderate image noise, label noise, and variations in scale or rotation, as they learn robust, hierarchical\footnote{CNNs operate by first detecting local, low-level features such as edges, corners, and textures in early layers. Deeper layers combine these primitives into higher-level features that capture complex shapes and semantic structure. This feature hierarchy allows CNNs to distinguish objects even under degraded imaging conditions, provided there is enough semantic variation in the data.} representations of visual features.


The readout images lack semantic structure and variability and differ fundamentally from natural-image datasets. The information needed for classification is localized to fixed, small patches around lattice sites, meaning that one of CNNs’ primary advantages—spatial invariance—offers little benefit. In fact, enforcing invariance can hinder performance by encouraging the model to search for features that are not physically meaningful in this highly constrained setting.



The challenge is amplified by the difficulty of collecting large, clean training datasets. Neutral atom readout is inherently slow, and acquiring images at scale is time-consuming. The process can also introduce label noise due to state preparation or measurement errors, further reducing the effectiveness of supervised training. As a result, models must learn from limited and imperfect data, unlike in typical vision benchmarks for which millions of labeled examples are available. These limitations become critical in a low signal-to-noise ratio (SNR) regime\footnote{The photon count distributions for bright ($\ket{1}$) and dark ($\ket{0}$) states overlap due to noise sources such as photon shot noise and camera readout noise.}, where the local intensity patterns that CNNs learn to detect the signal become statistically indistinguishable from random noise fluctuations. Consequently, the network struggles to maintain discrimination accuracy between qubit states.

\section{Enabling readout for low photon counts}



Neutral atom readout uses site-resolved fluorescence imaging: each atom sits at a known location in the image, and the brightness of the corresponding pixel tells us how many photons we collected during a short illumination period. If we make this readout shorter, we collect fewer photons. The useful signal then shrinks in direct proportion to the readout time, while the random noise shrinks more slowly (roughly with the square root of time). This makes it progressively harder to tell bright (occupied) sites from dark (empty) ones as we speed up the measurement, creating a fundamental trade-off between readout speed and accuracy.

Standard image-processing methods—such as Gaussian smoothing, median filtering, or Wiener filtering—work reasonably well when the images are only moderately noisy. They can reduce fluctuations while keeping the overall pattern intact. In the low-photon regime, however, the noise becomes comparable to the signal itself. These filters then tend to blur sharp features and distort the characteristic light pattern at each site, removing exactly the details needed to decide whether a site is occupied or empty. In this regime, we need denoising methods that can clean up the image while preserving these subtle, discriminative features so that downstream classifiers can still achieve high accuracy.

\subsection{Improving signal-to-noise ratio}



Generative approaches offer a more powerful alternative to conventional techniques by learning a direct mapping from low-SNR (short exposure) to high-SNR (long exposure) images. Using paired experimental datasets, these models can exploit the repetitive lattice structure of neutral atom arrays, preserve per-site correlations, and selectively enhance meaningful features while suppressing noise, framing the task as \textit{image-to-image translation problem}.

\begin{figure}[t]
\centering
\includegraphics[width=\columnwidth]{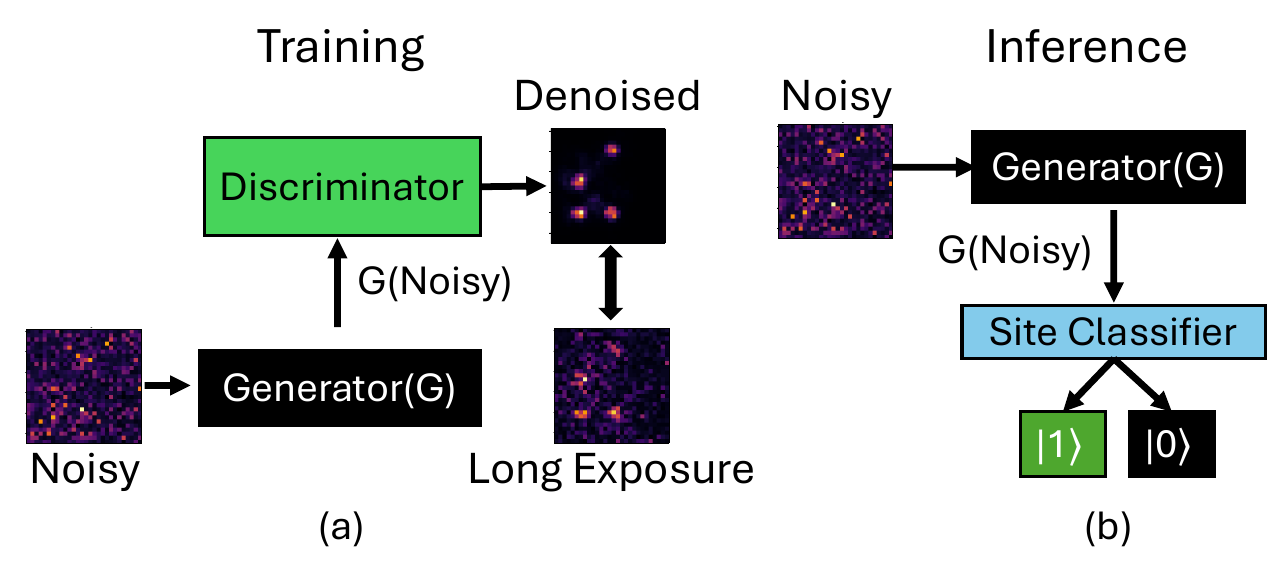}
\caption{(a) Training of Generative Adversarial Network~(GAN) with paired datasets of short exposure (Noisy) image and long exposure image and the discriminator tries to distinguish between the generated denoised image outputs from Generator(G) and the long exposure images while generator trains to fool the discriminator; (b) During inference, the learned generator generates denoised images and later sent to downstream readout classification.}
\label{fig:GAN_Training_Inference}
\end{figure}

Conditional generative adversarial networks (cGANs), such as Pix2Pix, are well-suited for image-to-image translation problems~\cite{pix2pix2017}. The generator restores missing spatial details, recovers contrast between bright (occupied) and dark (unoccupied) sites, and enhances fine-scale structure, while the discriminator acts as a critic to ensure that the generated outputs, the denoised images, are consistently close to the true high-photon-count counterparts as shown in Figure~\ref{fig:GAN_Training_Inference}(a). These models produce denoised images with the resolution and contrast required for accurate downstream classification in the low-photon regime as shown in Figure~\ref{fig:GAN_Training_Inference}(b).

\begin{figure}[t]
\centering
\includegraphics[width=\columnwidth]{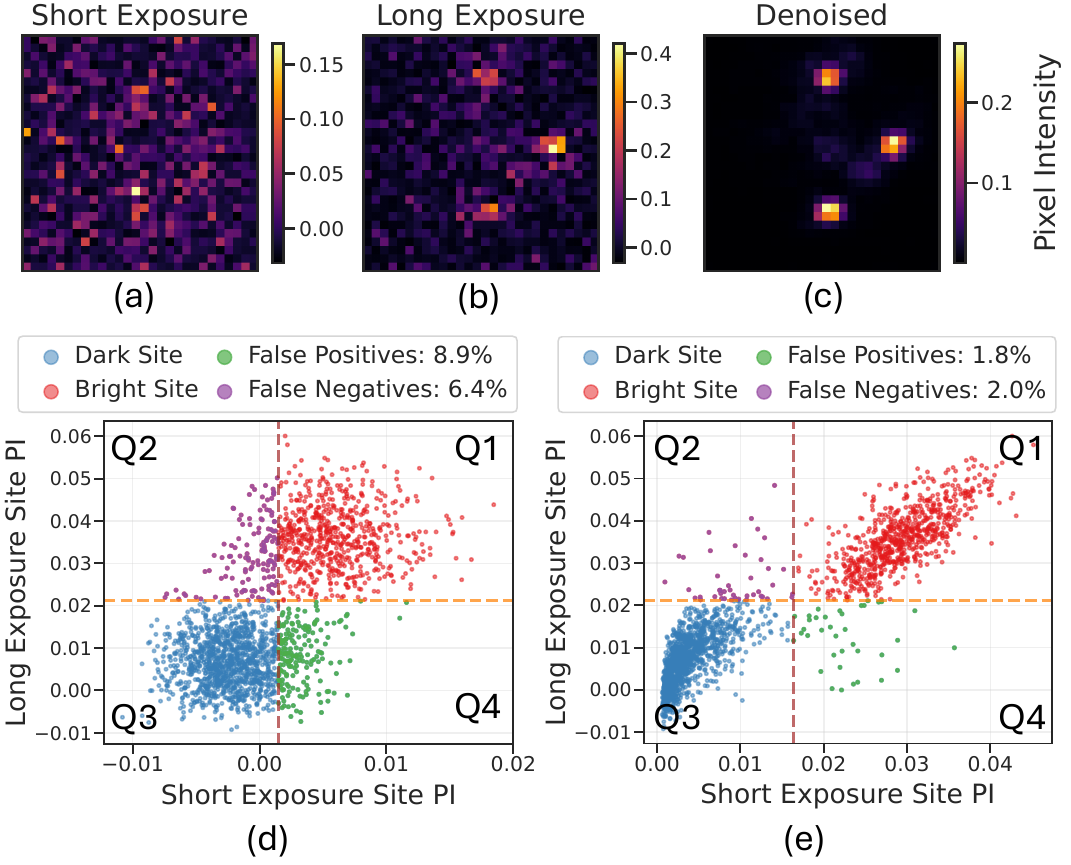}
\caption{Normalized average Pixel Intensity (PI) for the same experiment corresponding to (a) Signal attenuated path (short exposure) (b) Primary path (long exposure)  (c) Denoised image from short exposure image. We analyze plot average pixel intensity of (d) Short exposure, and (e) Denoised images, for the center site of all the images to evaluate the effect of denoising based on misclassifications of false positives in Quadrant Q4 and false negatives in Quadrant Q2, with $\sim\rm15\%$ without denoising and about $\sim\rm4\%$ with denoising.}
\label{fig:denoising_impact}
\end{figure}

Generative denoising improves low-SNR short-readout images by generating denoised images that are closer to the high-SNR long-exposure images by recovering much of the lost contrast. The noisy, low-SNR short exposure readout image in Figure~\ref{fig:denoising_impact}(a) is denoised to Figure~\ref{fig:denoising_impact}(c) that resembles the high-SNR long-exposure image in Figure~\ref{fig:denoising_impact}(b). 

To evaluate this quantitatively, Figure~\ref{fig:denoising_impact}(d - e) compares site-averaged intensities with the long-exposure reference for a particular site for a group of site-specific images. Ideally, we want to observe a diagonal line with minimal deviations.  Correctly denoised sites fall in Q1 (bright–bright) and Q3 (dark–dark), whereas Q2 (bright–dark) and Q4 (dark–bright) indicate mismatches. In Figure~\ref{fig:denoising_impact}(d), noise scatters many points into Q2 and Q4, but in Figure~\ref{fig:denoising_impact}(e) denoising pulls them back into Q1 and Q3, showing closer agreement with ground truth and fewer mismatches as expected. By improving effective SNR without altering the physical readout process, generative denoising directly enhances classification accuracy in low-photon regime, lowering logical error rates and reducing the error-correction overhead in large-scale quantum processors.


\subsection{Deep-dive into generative models for denoising}

While state-of-the-art generative models for image-to-image translation, such as Pix2Pix, have demonstrated impressive results in the natural-image domain, directly applying them to neutral atom readout is non-trivial. Natural-image datasets contain rich semantic diversity, i.e, edges, textures, and object boundaries, that models can exploit to learn generalizable mappings. In contrast, neutral atom images consist of highly repetitive, lattice-aligned point-spread patterns with relatively few distinguishing features beyond site intensity. This lack of semantic richness means that large, general-purpose models often allocate capacity to irrelevant structures, learn spurious correlations, or overfit to limited noise patterns. The challenge is compounded by the fact that dataset sizes in neutral atom experiments are orders of magnitude smaller than typical computer vision benchmarks, limiting the effectiveness of data-intensive GAN based denoising architectures.

In this setting, the goal of denoising is not to produce visually pleasing images in a perceptual sense but to maximize per-site classification fidelity. Each lattice site must be reconstructed as close as possible to its high-SNR counterpart, and spatial separation between neighboring sites must be preserved to avoid "blob merging" that can distort site-to-site correlation. Achieving this requires a model that emphasizes local structural accuracy and preserves the repetitive lattice geometry, rather than one optimized for global realism.

To meet these requirements, we developed a custom conditional GAN framework, \textbf{GANDALF (Generative Adversarial Network for Denoising At Low  Fluorescence)} as shown in Table~\ref{tab:gan_architecture}. We employ a similar structure as the Pix2Pix architecture, but tailor it to learn the mapping from low-SNR to high-SNR images of neutral atom readout. The generator follows an encoder–bottleneck–decoder design with skip connections and residual blocks, combining stability in training with strong feature preservation. Skip connections help mitigate vanishing gradients in deeper networks, allowing early-layer features to directly influence the final reconstruction, which is important for retaining high-frequency spatial detail that might otherwise be lost during downsampling. Residual blocks in the bottleneck further improve feature retention and stability by refining representations without disrupting the preserved information. 





Training optimizes a combination of adversarial loss, which encourages realism, and a heavily weighted pixel-wise L1 loss, which enforces fidelity to the high-SNR target at each pixel. Training is performed using the Adam optimizer with parameters $\beta_1 = 0.5$ and $\beta_2 = 0.999$, a learning rate of $0.0002$, batch size 16, and 30 training epochs.

The total generator loss is given by:
\begin{equation}
\mathcal{L}_{\text{G}} = \mathcal{L}_{\text{GAN}}(G, D) + \lambda_{L1} \, \mathbb{E}_{x, y} \left[ \| G(x) - y \|_{1} \right]
\end{equation}
where $x$ is the low-SNR input, $y$ is the high-SNR target, $\mathcal{L}_{\text{GAN}}$ measures the ability of generator to fool the discriminator, as mentioned in \cite{pix2pix2017}, and $\lambda_{L1}$ controls the trade-off between perceptual realism and pixel-level accuracy. In our experiments, $\lambda_{L1} = 200$ provided a good balance between structural preservation and noise suppression. To further improve structural preservation, the objective can be augmented with the Structural Similarity Index Measure (SSIM) and Peak Signal-to-Noise Ratio (PSNR) terms, which reward reconstructions that maintain perceptual structure and maximize SNR.

\begin{table}[t]
\centering
\caption{Generator and discriminator architectures}
\label{tab:gan_architecture}
\begin{tabular}{lcc}
\hline
\textbf{Layer} & \textbf{Filter shape} & \textbf{Output shape} \\
\hline
\multicolumn{3}{c}{\textbf{Generator}} \\
\hline
Input & N/A & $H \times W \times 1$ \\
Enc1 & $3 \times 3 \times 1 \times 64$ & $H \times W \times 64$ \\
Enc2 & $3 \times 3 \times 64 \times 128$ & $H/2 \times W/2 \times 128$ \\
Enc3 & $3 \times 3 \times 128 \times 256$ & $H/4 \times W/4 \times 256$ \\
Res Block 1 & $3 \times 3 \times 256 \times 256$ & $H/4 \times W/4 \times 256$ \\
Res Block 2 & $3 \times 3 \times 256 \times 256$ & $H/4 \times W/4 \times 256$ \\
Res Block 3 & $3 \times 3 \times 256 \times 256$ & $H/4 \times W/4 \times 256$ \\
Dec3 & $4 \times 4 \times 256 \times 128$  & $H/2 \times W/2 \times 128$ \\
Dec2 & $4 \times 4 \times 256 \times 64$  & $H \times W \times 64$ \\
Dec1 & $3 \times 3 \times 128 \times 1$ & $H \times W \times 1$ \\
\hline
\multicolumn{3}{c}{\textbf{Discriminator}} \\
\hline
Input & N/A & $H \times W \times 1$ \\
Conv1 & $3 \times 3 \times 1 \times 64$ & $H/2 \times W/2 \times 64$ \\
Conv2 & $3 \times 3 \times 64 \times 128$ & $H/4 \times W/4 \times 128$ \\
Conv3 & $3 \times 3 \times 128 \times 256$ & $H/8 \times W/8 \times 256$ \\
Conv4 & $3 \times 3 \times 256 \times 512$ & $H/16 \times W/16 \times 512$ \\
Output & $1 \times 1 \times 512 \times 1$ & $1 \times 1 \times 1$ \\
\hline
\end{tabular}
\end{table}

\subsection{Denoising limitations for neutral atom readout}

In the low-SNR regime and with low structural variability in readout images, GAN-based denoising can introduce systematic errors despite producing visually cleaner images. Overfitting to recurring noise patterns may distort true signal distributions, and in borderline cases where occupancy depends on subtle intensity differences, the generator may hallucinate sites or suppress faint signals, reducing reconstruction fidelity.

\subsubsection{Mode collapse and data limitations}

A common failure mode for generative models is mode collapse, where the generator outputs uniformly “bright” or “dark” patterns that satisfy the discriminator but fail to preserve true site occupancy. The repetitive lattice geometry and limited calibration datasets in atom readout increases chances, leading to memorization of noise or spurious correlations that introduce artificial features or suppress real signals. These errors directly degrade downstream classification, especially under unseen noise conditions.

\subsection{Training strategies to mitigate failure modes}

Mode collapse can be reduced by preventing the discriminator from becoming overconfident through label smoothing, i.e, assigning real samples a label of 0.9 instead of 1.0 and fake samples 0.1 instead of 0.0. We further reduce collapse and hallucination risk by updating the discriminator less frequently than the generator, giving the generator more opportunities to refine outputs without being prematurely constrained.

Dropout in the discriminator encourages focus on robust, generalizable features rather than memorizing noise patterns. To promote smooth convergence and reduce oscillatory dynamics, cosine annealing learning rate schedules are employed. Finally, early stopping based on validation L1 loss captures the best model before it begins overfitting to training-set noise characteristics. Together, these measures allow the model to reliably reconstruct from low-SNR to high-SNR readouts without introducing artificial features, improving classification accuracy in the low-photon-count regime.

\subsection{Downstream Readout Classification Task}

Accurate readout classification is critical for high-fidelity state detection. Traditional thresholding approaches are lightweight but fragile in low-photon regimes.  Matched filtering offers interpretability but similarly degrades at low SNR. Convolutional Neural Networks~(CNN)-based classifiers achieve higher accuracy at the cost of excessive complexity with $\sim74$ million parameters for a $3\times 3$ atom array \cite{phuttitarn2023enhanced}, making it impractical to scale them for larger atom arrays. 
\begin{figure}[t]
    \centering
    \includegraphics[width=\linewidth]{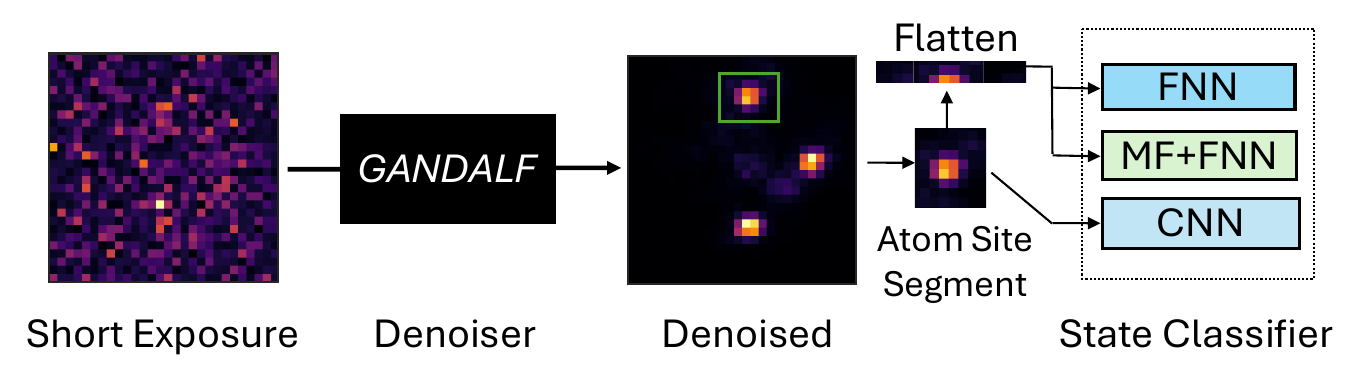}
    \caption{Overview of Readout Classification with \oursnspace. The short exposure noisy image with low signal-to-noise ratio (SNR) is denoised to obtain high quality, high SNR denoised image. Depending on the constraints on the expected input of the classifier, the denoised image is further segmented at each of the atom sites and/or flattened and then feed into the state classification methods like Feed Forward Neural Networks~(FNN), Matched Filter with Neural Network~(MF+NN) and Convolutional Neural Networks~(CNN).}
    \label{fig:readout_classification}
\end{figure}

\ours mitigates this challenge by denoising short-exposure measurements, reducing classification hardness and enabling the use of lightweight models without sacrificing fidelity. The task of neutral atom readout is formulated as binary classification. Given a fixed-size site-centered patch, determine whether the atom is bright-occupied ($\ket{1}$) or dark-unoccupied ($\ket{0}$). Each site-specific model is trained per lattice location to account for spatial variations in signal and noise.  

\begin{figure}[h]
    \centering

    \includegraphics[width=0.95\linewidth]{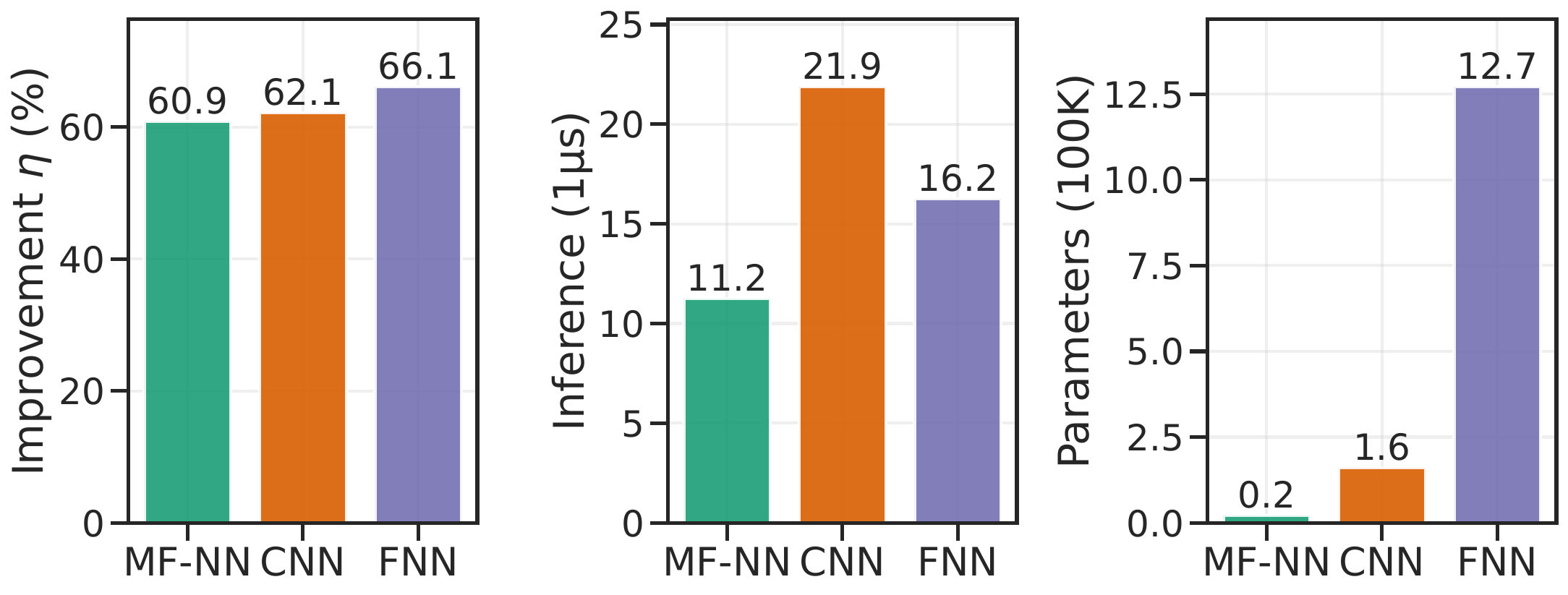}
    \caption{Comparison of readout classification methods, including, Matched Filter with Neural Networks (MF-NN), Convolutional Neural Networks (CNN) and Feed Forward Neural Networks (FNN). We report the metrics: relative infidelity reduction ($\eta$) compared to the baseline CNN~(Site), inference time and total model parameters.
    }
    \label{fig:readout_classification_comparison}
\end{figure}

We design and evaluate matched filter with neural network (MF+NN), feedforward neural network (FNN), and 2D-CNN classifiers, to compare their strengths and limitations. Figure~\ref{fig:readout_classification} illustrates the classification pipeline. The process begins with denoising the readout image with \oursnspace, followed by extraction of fixed-size patches centered on each lattice site. CNNs operate directly on these patches, whereas FNN and MF+NN use flattened patches. In the case of MF+NN, matched-filter features are additionally incorporated.



Classifier improvement is measured using relative infidelity reduction $\eta = 1 - \delta_{\text{method}}/\delta_{\text{CNN(Site)}}$, where $\delta$ is classification inaccuracy, and higher $\eta$ implies better inaccuracy reduction. As shown in Fig.~\ref{fig:readout_classification_comparison}, all classifiers benefit substantially from \oursnspace, achieving over 60\% reduction in inaccuracy relative to the baseline. At the shortest readout durations, error is reduced by $2.5\times$, while model size and inference latency decrease by up to $16.5\times$ and $5\times$ respectively, compared to the baseline. Similarly, end-to-end latency including \ours is $4\times$ faster than the CNN~(Array) baseline.  

Among site-specific classifiers, FNN yields the highest accuracy gains, while MF+NN matches this performance with fewer parameters by exploiting matched-filter features, both surpassing baseline CNNs. Together with \oursnspace, they deliver scalable, low-latency readout well-suited for fault-tolerant neutral atom architectures.

\section{Scalable and Fast Readout for QEC}



Supporting QEC at scale requires high-fidelity in-circuit measurements across hundreds or thousands or even more qubits. At such scales, readout speed is a primary bottleneck. While two-qubit gates finish in less than a microsecond, the current readout can take tens of milliseconds. To overcome this gap, two factors are critical:
(1) Image acquisition — the duration needed to collect scattered photons. (2) Classification latency — the time required to process the signal and infer qubit states for thousands of qubits. Shorter image acquisition times reduce delay but make classification harder, and often demand more sophisticated models with higher inference costs.  In this section, we analyze the speed–accuracy tradeoff and the scalability of GANDALF.


\subsection{Scaling to larger atom arrays}

As neutral atom systems scale to thousands of qubits, readout fidelity must remain robust across increasingly large arrays. A major challenge is that many denoising and classification approaches are tailored to fixed input dimensions, making them impractical for scaling without retraining or re-engineering. This creates concerns about computational load and consistency of performance at scale.

Our approach overcomes this by employing a fully convolutional denoising network, which naturally accepts images of arbitrary size. Convolutional filters operate locally and reuse the same learned kernels across space, allowing the model to generalize seamlessly from small calibration arrays to much larger lattices without modifying architecture or parameters. To evaluate scalability, we test GANDALF on synthetically generated large atom arrays by stitching single-atom images.


Table~\ref{tab:denoising_scaling} reports Peak Signal-to-Noise Ratio (PSNR), Structural Similarity Index Measure (SSIM), and mean L1 loss for denoising as atom arrays scale from $4 \times 4$ to $64 \times 64$. The results show only slight degradation with increasing grid size, i.e, PSNR drops by less than 0.5dB, SSIM decreases by 0.02, and mean L1 loss remains nearly unchanged. With fixed model complexity, the computational cost of \ours grows marginally with image size, keeping per-site inference time nearly constant.

\begin{figure}[ht]
    \centering
    \includegraphics[width=\linewidth]{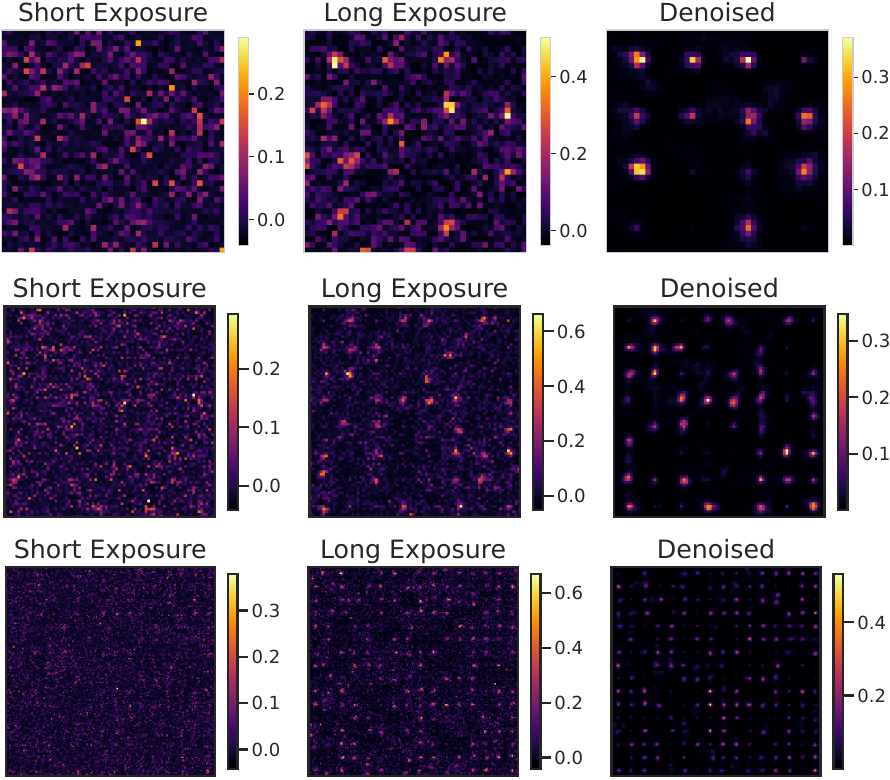}
    \caption{Scalability of \ours with consistent recovery of site contrast for atom arrays of $n\times n$ atoms for n = 4,8,16.}
    \label{fig:scaling_denoising}
\end{figure}

\begin{table}[b]
    \centering
    \caption{Scalability of denoising across atom array sizes.}
    \vspace{-0.05in}
    \label{tab:denoising_scaling}
    \begin{tabular}{c|c|c|c|c}
        \hline
        \multirow{2}{*}{Metric} & \multicolumn{4}{c}{Atom array grid size ($n\times n$)} \\
        \cline{2-5}
        & $4 \times 4$ & $8 \times 8$ & $32 \times 32$ & $64 \times 64$ \\
        \hline
        PSNR (dB)    & 25.17& 25.06 & 24.79 & 24.86 \\
        SSIM         & 0.3337 & 0.3368 &  0.3134 & 0.3187 \\
        Mean L1 loss & 0.0362 & 0.0365 &  0.03701 & 0.0367 \\
        Latency($ms$) & 0.201 & 0.514 & 9.091 & 33.490  \\
        \hline
    \end{tabular}
\end{table}

Figure~\ref{fig:scaling_denoising} shows the denoiser trained on $3\times3$ arrays generalizes to larger arrays upto $16\times 16$, while preserving bright–dark separation and site contrast. This demonstrates \ours trained on small arrays can transfer directly to large-scale neutral atom processors with thousands of atoms.

\subsection{Denoising Latency and Practical Scalability}

We evaluate the \ours generator (4.7 million parameters) across grid sizes, batch sizes, and parallel executions to assess its suitability for real-time QEC pipelines. Although input size affects inference time, the fully convolutional design ensures sublinear growth on a single GPU, with latency remaining within tens of ms even for $32\times32$ atom arrays as shown in Table~\ref{tab:denoising_scaling}. These trends indicate that denoising is practical at scale, with GPUs or other accelerators.  




\begin{table}[h]
    \centering
    \small
    \caption{Denoising latency and throughput (TP) for $3\times3$ atom array with the batch sizes on NVIDIA\textsuperscript{\textregistered} GeForce RTX\texttrademark~2080~Ti GPU.}
    \vspace{-0.05in}
\label{tab:batch_scaling}
    \begin{tabular}{c|c|c|c|c|c}
        \hline
        \multirow{2}{*}{Metric} & \multicolumn{5}{c}{Batch Size} \\
        \cline{2-6}
        & 1 & 8 & 32 & 64 & 128 \\
        \hline
        TP (img/s) & 460.0 & 3330.3 & 9225.0 & 11721.4 & 12711.4 \\
        Latency ($\rm ms$) & 2.174 & 0.300 & 0.108 & 0.085 & 0.079 \\
        \hline
    \end{tabular}
\end{table}

\begin{table}[h]
\vspace{-0.10in}
    \centering
    \small
    \caption{Denoising latency and throughput (TP) for a $3\times3$ atom array (batch size 32) using parallel models on NVIDIA\textsuperscript{\textregistered} GeForce RTX\texttrademark~2080~Ti GPU.}
    \label{tab:parallel_scaling}
    \begin{tabular}{c|c|c|c|c}
        \hline
         \multirow{2}{*}{Metric} & \multicolumn{4}{c}{Parallel Models} \\
        \cline{2-5}
       & 4 & 8 & 16 & 32 \\
        \hline
        TP/model (img/s) & 10615.5 & 11204.0 & 11786.8 & 12083.5 \\
        Latency $(ms)$ & 0.094 & 0.089 & 0.085 & 0.083\\
        
        \hline
    \end{tabular}
\end{table}

Throughput can be further improved using batching and parallelism. With batching, the GPU processes multiple images simultaneously, reducing per-image latency from $2.174$ ms to $0.079$ ms at a batch size of 128, as shown in Table~\ref{tab:batch_scaling}, for an atom array of $3\times3$, similar to the dataset used~\cite{phuttitarn2023enhanced}. 

Parallel execution provides another dimension of scaling, as shown in Table~\ref{tab:parallel_scaling}, throughput per model remains nearly constant with latencies of 80–90 $\mu$s, while memory usage increases roughly linearly with the number of models. In practice, moderate batching combined with 8–32 concurrent models keeps denoising well within QEC timing budgets while sustaining high throughput. We envision further reductions in denoising latency by using more powerful GPUs, as our GPU setup has modest computational capabilities and 5x lower memory bandwidth than NVIDIA\textsuperscript{\textregistered} H100 GPUs.

\subsection{Pipelining classification across cycles}

Readout, from image acquisition to state classification, must operate within the strict timing budget of QEC cycles. State preparation and rearrangement is performed to assemble defect-free atom arrays, before the start of error correction. Later, QEC proceeds through repeated cycles of gate execution, and measurement.
As summarized in Table~\ref{tab:qec_timing}, gate operations are fast, but readout remains the dominant per-cycle bottleneck due to millisecond-scale image acquisition and the additional classification step. 

\begin{table}[b]
\centering
\small
\caption{Typical QEC cycle latency for neutral-atom systems.}
\vspace{-0.05in}
\label{tab:qec_timing}
\begin{tabular}{l c}
\toprule
\textbf{Process} & \textbf{Typical Duration} \\
\midrule
State preparation (one-time) & $\sim 100$ ms \\
Two-qubit gate operation & few $\mu$s \\
Image acquisition & few ms \\
Classification & few $100~\mu$s \\
\bottomrule
\end{tabular}
\end{table}



\begin{figure}[!t]
    \centering
    \includegraphics[width=0.8\linewidth]{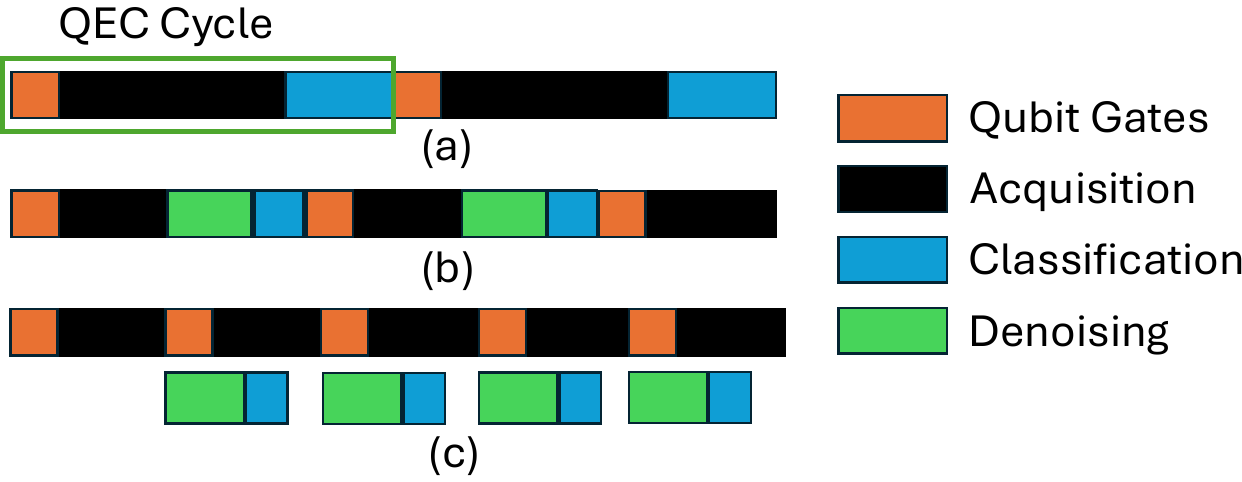}
    \caption{Overview of qubit readout modes: (a) Baseline – sequential, no GANDALF, longer acquisition. (b) Unpipelined: sequential with GANDALF, shorter acquisition but added denoising latency. (c) Pipelined: overlaps classical processing with other stages, hiding denoising and classification latency.}
    \label{fig:readout_pipeline}
\end{figure}

In current approaches, classification is performed immediately after image acquisition, with inference latency accumulating at every QEC cycle as in Figure~\ref{fig:readout_pipeline}(a) and without GANDALF, the acquisition time is significantly larger compared to when denoising is enabled as shown in Figure~\ref{fig:readout_pipeline}(b). While the sequential execution overhead is negligible for small systems, it becomes significant as arrays scale to hundreds of qubits. Baseline methods typically employ both single-site and multi-site CNN classifiers with classification taking up to $\sim50~\mu$s per site, and $500~\mu$s for a $3\times3$ array, respectively,  almost an order of magnitude slower for multi-site. Similar scaling to larger arrays quickly drives classification into the multi-millisecond regime, creating a major throughput bottleneck.

Our approach mitigates this by relying on lightweight single-site classifiers such as eedforward neural networks and  matched filter with neural networks~(MF+NN) that run in parallel across sites and achieve up to $5\times$ faster inference than CNN~(Array). Multi-site correlations are instead captured in the denoising stage, allowing classification to remain compact and scalable. Moreover, we propose pipelining classification with image acquisition to effectively hide its latency, eliminating it as a limiting factor for large-scale QEC.


In existing small-scale neutral atom readout systems, classification is performed immediately after image acquisition to trigger reset and atom reloading, making it a per-cycle bottleneck. Recent works suggest that reset\cite{Geher2025} and atom loss\cite{baranes2025leveragingatomlosserrors} can be tolerated. As the chance of atom loss is rare and does not introduce spurious correlations, reloading can instead be deferred by one round, allowing classification to overlap with the next measurement as shown in Fig.~\ref{fig:readout_pipeline}(b). This reduces idle time and mitigates idling-induced errors. 




With an unpipelined design, the time to execute $d$ rounds of syndrome extraction for QEC is,
\begin{equation}
    T_{\text{QEC}}^{\text{unpipelined}} = d \times (t_{\text{gate}} + t_{\text{readout}} + t_{\text{classification}}),
\end{equation}

In our scheme, the effective latency becomes  
\begin{equation}
    T_{\text{QEC}}^{\text{pipelined}} = d \times ( t_{\text{readout}} + t_{\text{gate}} )+(t_{\text{denoise}} + t_{\text{classification}}),
\end{equation}

With pipelining, classification cost is incurred only after the final round, with most of its cost hidden behind acquisition. Gains are largest at short readout times (1.5\,ms). Crucially, classification never becomes the limiting factor, ensuring QEC is constrained mainly by acquisition time.

\section{Methodology}

\noindent\textbf{Quantum hardware and dataset.} We obtained datasets with readout images from the atom array of $3\times3$ with 9-qubit neutral atom system as mentioned in~\cite{phuttitarn2023enhanced} with two different spacings between atoms of $5~\mu m$ and $9~\mu m$ resulting in $28\times28$ and $32\times32$ pixel images respectively. The qubit atom array is illuminated for a variable readout duration and the scattered photons are collected using an EMCCD camera. A total of 14 different readout durations ranging from $15~\rm ms$ to $100~\rm ms$ are used to obtain $27000$ single-site long-exposure images for each duration. For each long exposure image, the dataset has corresponding short-exposure images obtained by attenuating the signal by $10\times$, resulting in effective readout durations ranging from $1.5~\rm ms$ to $10~\rm ms$.


\noindent\textbf{Training Procedure.} Prior to training, all readout images are normalized to ensure consistent scaling, numerical stability, and faster convergence. Each image $I_{ij}$ is zero-centered and normalized according to $I_{ij}' = \frac{I_{ij} - \mu}{I_{\text{max}} - I_{\text{min}}}$, where $\mu$, $I_{\text{max}}$ and $I_{\text{min}}$ are the mean, maximum and minimum pixel intensities in the training subset, respectively. \ours uses paired long-exposure and short-exposure images with 65-15-20\% as train-validation-test split. The classification models are trained for individual sites in the atom array and uses corresponding site patches of the readout image.

\noindent\textbf{Simulation framework.} The effect of readout errors on logical error rates of surface codes and BB codes is evaluated using the Stim\cite{gidney2021stim} stabilizer simulator with PyMatching and BP-OSD-CS10 as the respective decoders. 


\noindent\textbf{Hardware.} We evaluate our models on NVIDIA\textsuperscript{\textregistered} GeForce RTX\texttrademark~2080~Ti GPU and Intel\textsuperscript{\textregistered} Xeon\texttrademark~Silver~4216 CPU.

\noindent\textbf{Software.} All readout state discriminators are implemented in Python. We use PyTorch framework to build, train, and test all the neural network models in our proposed design.

\section{Evaluations}

\textbf{Evaluation metrics.} We evaluate our method, GANDALF, at three levels: (1) \textit{State Classification:} averaged-site inaccuracy and relative infidelity reduction ($\eta$) over the baseline, including post-selection filtering to capture accuracy–latency tradeoffs. (2) \textit{System level:} Logical Error Rate~(LER) for surface code and Bivariate Bicycle~(BB) code, accounting for idling errors. (3) \textit{QEC cycle latency:} execution-time reduction of repeated syndrome extraction for QEC codes.

\begin{figure}[h]
    \centering
    \includegraphics[width=\linewidth]{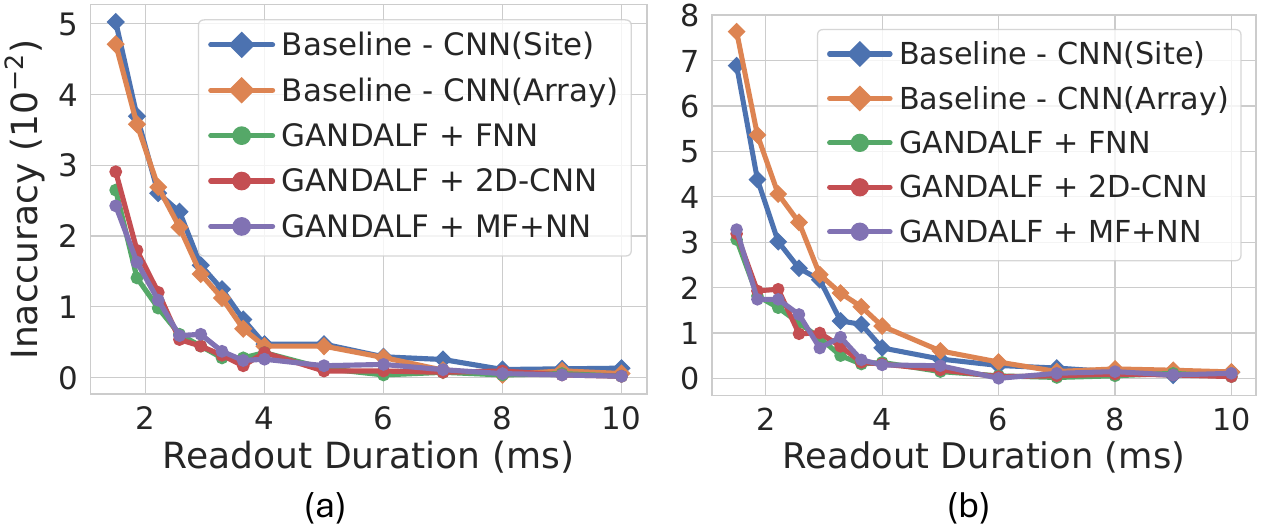}
    \caption{Readout classification inaccuracy of baseline CNN architectures~\cite{phuttitarn2023enhanced}, CNN~(Site) and CNN~(Array) compared to GANDALF with fast and small classifiers based on feed forward neural network (FNN), CNN and Matched Filter with neural network (MF+NN) for the atom spacing of (a) $5~\mu\rm m$ and (b) $9~\mu\rm m$ over all the readout duration ranging from $1.5~\rm ms$ to $10~\rm ms$.}
    \label{fig:readout_inaccuracy_vs_readout_duration}
\end{figure}

\subsection{Impact on readout accuracy and readout duration}

Figure~\ref{fig:readout_inaccuracy_vs_readout_duration} shows the improvement in readout inaccuracy as a function of readout duration for both $5~\mu$m and $9~\mu$m atom spacings. \ours delivers substantial gains at short exposure times, where SNR is lowest. Our method achieves $2.8\times$ reduction in readout error compared to the baseline CNN(Site) at $1.5~\rm ms$. Furthermore, our method sustains below 1\% inaccuracy up to $2.9~\rm ms$ for $5~\mu$m and $3.2~\rm ms$ for $9~\mu$m, compared to $3.6~\rm ms$ and $4~\rm ms$ for the baseline, respectively. On average, relative infidelity factor, i.e, $\eta = 1 - \delta_{\text{method}}/\delta_{\text{CNN(Site)}}$, where $\delta$ denotes the classification inaccuracy, improves by $\sim$64\% for atom spacing $5~\mu$m and $\sim$53\% for the $9~\mu$m lattices.

\begin{figure}[t]
    \centering
    \includegraphics[width=0.95\linewidth]{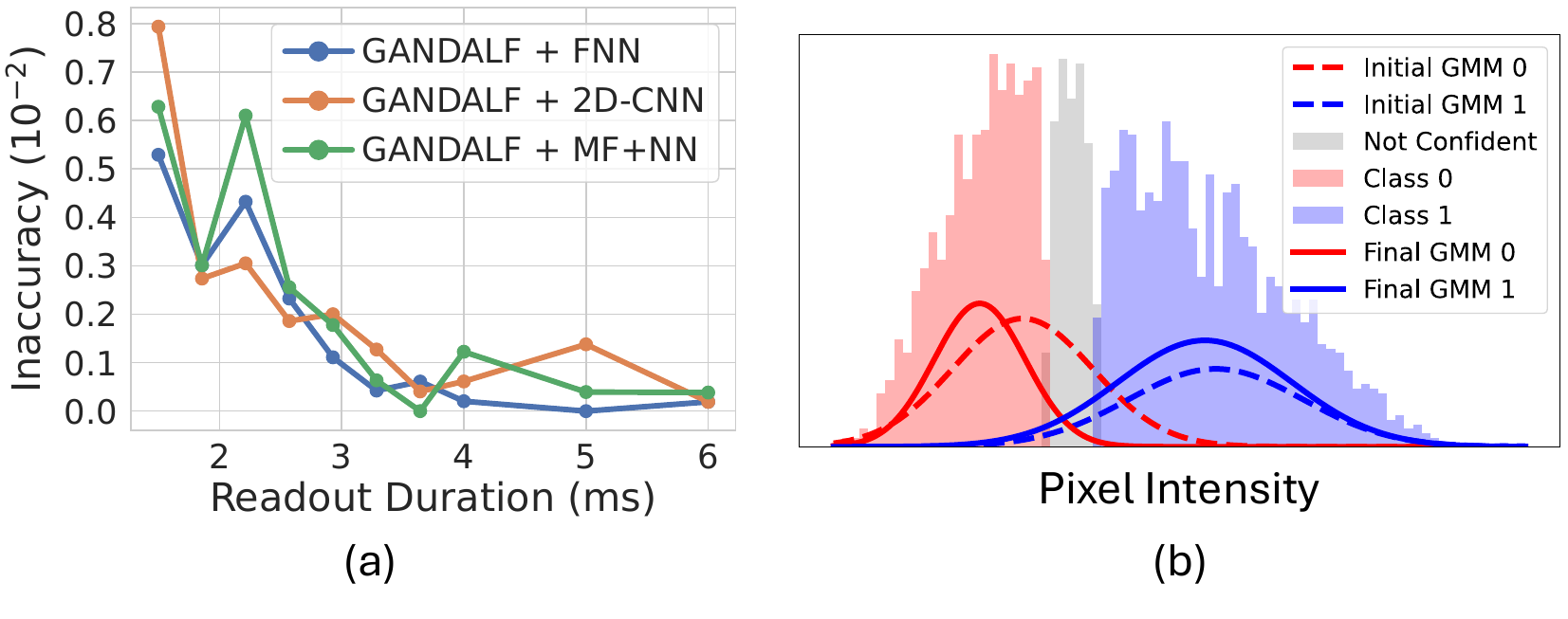}
    \caption{(a) Readout classification inaccuracy for post-selected dataset to avoid the overlapping, less-confident region over readout durations ranging from $1.5~\rm ms$ to $6~\rm ms$ for GANDALF with classifiers based on the feed forward neural network (FNN), CNN and Matched Filter with neural network (MF+NN). (b) The overlap of the histograms of pixel intensity of the readout image depicts the low-confidence images. The post-selection removes low-confidence images using the fitted Gaussian Mixture Models (GMMs).}
    \label{fig:post_selected_5um}
\end{figure}

The dataset used in this work is imperfect, but higher-quality data is expected in the near future. To estimate the best-case improvements achievable with denoising, we construct a confidence-filtered dataset using Gaussian Mixture Model (GMM)–based confidence filtering. Instead of relying solely on raw classifier outputs, we fit a two-component GMM to the distribution of decision scores, separating high-confidence bright and dark populations from ambiguous cases as shown in Figure~\ref{fig:post_selected_5um}(b). By post-selecting only the confident subsets, we maintain sub-1\% readout inaccuracy even at the shortest duration of $1.5~\rm ms$ as shown in Figure~\ref{fig:post_selected_5um}(a). This demonstrates the potential to reduce readout times close to sub-millisecond regime for Cs arrays aiding \ours with
confidence-aware post-selection.

\subsection{Logical error rate}
To quantify the end-to-end impact of readout fidelity and latency on fault-tolerant performance, we evaluate logical error rates (LER) using both bivariate bicycle (BB) codes and surface codes (SC). Idling errors during measurement were modeled through the Pauli twirling approximation~\cite{Tomita2014} characterized by $T_1$ and $T_2$ times, allowing us to capture the tradeoff between longer readout times with improved accuracy but increased idling error against shorter readout times with reduced idling error but lower accuracy.

\begin{figure}[!t]
    \centering
    \includegraphics[width=0.91\linewidth]{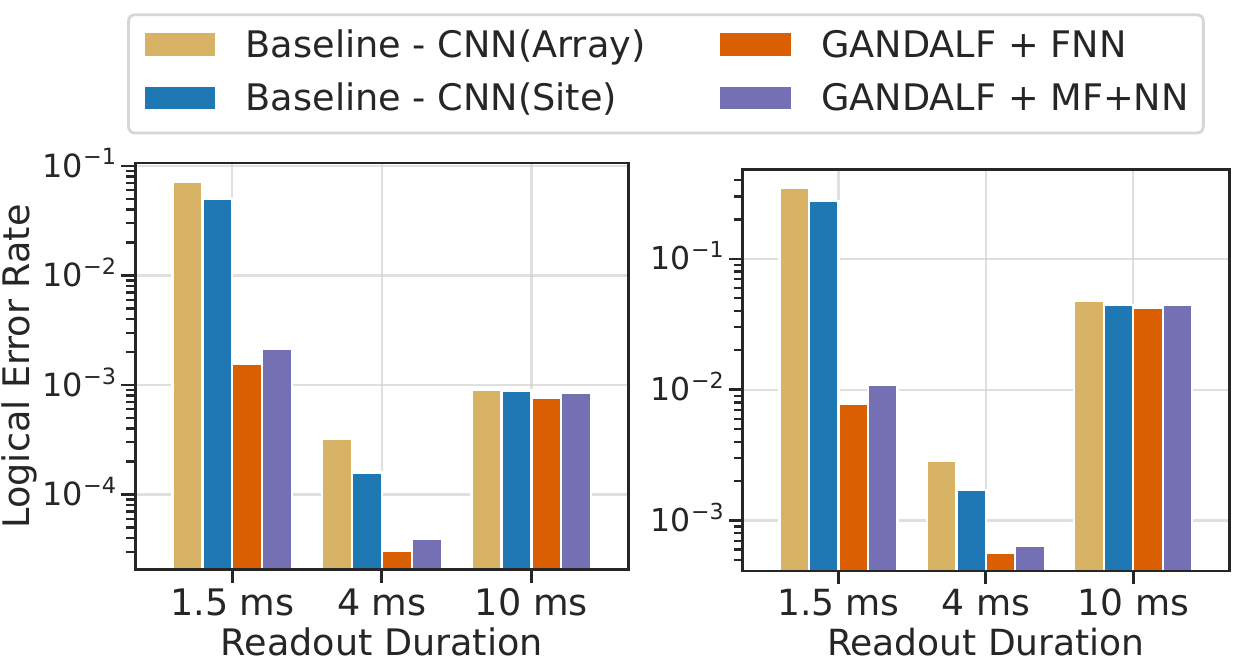}
    \caption{Logical Error Rate (LER) comparison of surface $d=11$ (left) and Bivariate Bicycle (BB) code $d=12$ for the readout durations of $1.5~\rm ms$, $4~\rm ms$ and $10~\rm ms$ incorporating the effect of idling errors.}
    \label{fig:9um_LER_eval}
\end{figure}

Figure~\ref{fig:9um_LER_eval} indicates an intermediate readout duration of approximately 4ms achieves the best overall LER among the readout durations for $9~\mu\rm m$ spacing, similar trends are observed for $5~\mu\rm m$ spacing. \oursnspace-enhanced FNN achieves $5\times$ reduction in LER compared to the CNN(Site) baseline at $4~\rm ms$ and upto $35.8\times$ reduction at $1.5~\rm ms$ for BB codes, highlighting the benefits of better readout accuracy at shorter readout duration.  

At longer readout times, the gains from higher classification accuracy are outweighed by accumulated idling errors. Conversely, at shorter readout times, insufficient accuracy dominates, leading to higher LER. These results highlight the fundamental tradeoff between readout duration, readout accuracy, and LER, emphasizing the need for optimized readout strategies that balance these competing effects in large-scale neutral atom architectures.


\begin{figure}[ht]
    \centering
    \includegraphics[width=0.95\linewidth]{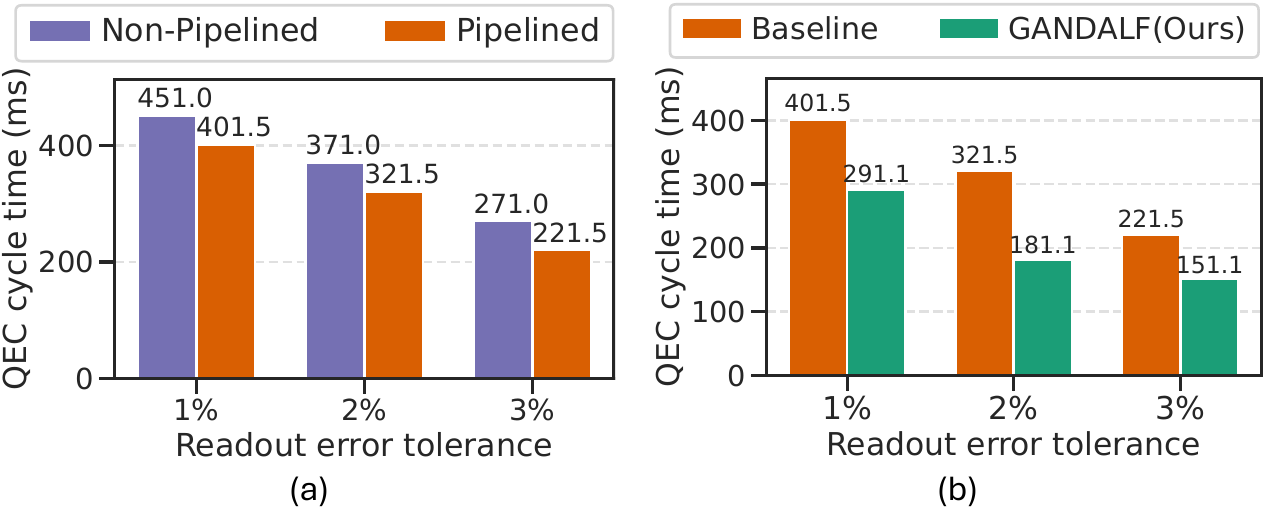}
    \caption{Total QEC Execution time for run for 100 rounds readout error tolerance of 1\%, 2\% and 3\%. (a) Baseline (without denoising) with and without pipelining (b) Pipelined  baseline design with compared to \ours (ours).}
    \label{fig:qec_timing_cycle_compare}
\end{figure}

\subsection{Reducing QEC cycle latency}  

Pipelining overlaps denoising and classification with subsequent acquisitions, so their cost is paid only once at the final round. As shown in Fig.~\ref{fig:qec_timing_cycle_compare}, this reduces QEC execution time by over 10\% ($50~\rm ms$) across 100 rounds, with the largest gains at short readout times, where classification dominates the cost. Incorporating \ours further reduces QEC cycle time, achieving up to $1.77\times$ reduction in the pipelined baseline. 

At the shortest readout time of $1.5~\rm ms$, the execution time for 100 rounds decreases from $0.19~\rm s$ with the baseline to $0.152~\rm s$ with \ours and pipelining. The advantage grows with number of QEC rounds as conventional pipelines  adds constant classification latency per round. While the overlapping pipelined scheme removes this latency from critical path.

\section{Conclusions}


We present GANDALF, a system architecture that optimizes neutral atom readout by addressing fidelity and latency, the main bottlenecks in scalable quantum error correction. Our framework uses explicit denoising with image translation to reconstruct clear signals from short, low-photon measurements, enabling reliable classification at up to $1.6\times$ shorter readout times. Combined with lightweight classifiers and a pipelined readout design, GANDALF reduces the logical error rate up to $35\times$ for BB code and up to $5\times$ for Surface code and enables  $1.77\times$ lower QEC cycle time compared to state-of-the-art CNN-based readout for neutral atom arrays. Fully convolutional networks used in our design enable scaling to very large arrays, with batching and parallelism achieving millisecond-level throughput, together establishing a practical path to high-fidelity, low-latency, and scalable readout for large-scale neutral atom processors.

\bibliographystyle{ACM-Reference-Format}
\bibliography{refs}


\end{document}